\pdfoutput=1
\documentclass[10pt,twocolumn,letterpaper]{article}

\usepackage{text}             
\usepackage[table]{xcolor}
\usepackage{wrapfig}

\usepackage{xcolor}
\usepackage{soul}

\usepackage{graphicx}
\usepackage{booktabs}
\usepackage{bbm}
\usepackage{nicefrac}
\usepackage{empheq}
\usepackage[normalem]{ulem}
\usepackage{wrapfig}
\usepackage{multirow}
\usepackage{stmaryrd}
\usepackage{bm}
\usepackage{subcaption}
\usepackage{pifont}
\usepackage[table]{xcolor}
\usepackage{placeins}
\usepackage{listings}
\definecolor{cvprblue}{rgb}{0.21,0.49,0.74}
\newcommand{\datasetname}{\textit{OrigamiCode}}
\newcommand{\papertitle}{\textit{Learn2Fold}}

\usepackage{algorithm}
\usepackage{algpseudocode}
\usepackage{algorithmicx}
\algdef{SE}[DOWHILE]{Do}{doWhile}{\algorithmicdo}[1]{\algorithmicwhile\ #1}

\usepackage{amsmath,amssymb}

\usepackage[pagebackref,breaklinks,colorlinks,allcolors=cvprblue]{hyperref}

\newcommand\blfootnote[1]{%
  \begingroup
  \renewcommand\thefootnote{}\footnote{#1}%
  \addtocounter{footnote}{-1}%
  \endgroup
}

\newcommand{\TopOne}[1]{\cellcolor{blue!35}{#1}}
\newcommand{\TopTwo}[1]{\cellcolor{blue!20}{#1}}

\newcommand{\GOne}[1]{\cellcolor{orange!35}{#1}}
\newcommand{\GTwo}[1]{\cellcolor{orange!20}{#1}}

\title{Learn2Fold: Structured Origami Generation with World Model Planning}

\author{
Yanjia Huang$^{1,2*}$  \quad
Yunuo Chen$^{1*}$ \quad 
Ying Jiang$^{1*}$ \quad 
Jinru Han$^{1}$ \quad \\
Zhengzhong Tu$^{2}$ \quad 
Yin Yang$^{3}$ \quad
Chenfanfu Jiang$^{1}$
}

\begin{document}

\setcounter{footnote}{0}

\twocolumn[{%
\renewcommand\twocolumn[1][]{#1}%
\maketitle

\begin{center}
    \centering
    \captionsetup{type=figure}
    \includegraphics[width=\textwidth]{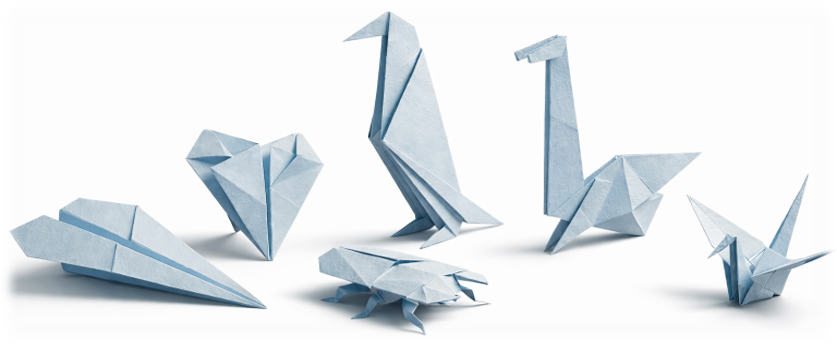}
  \caption{\textbf{Teaser.} From simple planes to complex articulated forms, Learn2Fold plans origami folding sequences that respect geometric constraints and anticipate future consequences, enabling robust generalization across unseen crease patterns.}
  \label{fig:teaser}
\end{center}
}]
\blfootnote{* Equal contribution. $^{1}$ UCLA, 
$^{2}$ Texas A\&M University, 
$^{3}$ University of Utah. 
\url{yanjia_0812@tamu.edu}, 
\url{yunuoch@math.ucla.edu}, 
\url{anajymua@gmail.com}, 
\url{jinruhan1219@g.ucla.edu}, 
\url{tzz@tamu.edu}, 
\url{yin.yang@utah.edu}, 
\url{cffjiang@ucla.edu}}

\begin{abstract}
    The ability to transform a flat sheet into a complex three-dimensional structure is a fundamental test of physical intelligence. Unlike cloth manipulation, origami is governed by strict geometric axioms and hard kinematic constraints, where a single invalid crease or collision can invalidate the entire folding sequence. As a result, origami demands long-horizon constructive reasoning that jointly satisfies precise physical laws and high-level semantic intent. Existing approaches fall into two disjoint paradigms: optimization-based methods enforce physical validity but require dense, precisely specified inputs, making them unsuitable for sparse natural language descriptions, while generative foundation models excel at semantic and perceptual synthesis yet fail to produce long-horizon, physics-consistent folding processes. Consequently, generating valid origami folding sequences directly from text remains an open challenge. To address this gap, we introduce \textbf{Learn2Fold}, a neuro-symbolic framework that formulates origami folding as conditional program induction over a crease-pattern graph. Our key insight is to decouple semantic proposal from physical verification. A large language model generates candidate folding programs from abstract text prompts, while a learned graph-structured world model serves as a differentiable surrogate simulator that predicts physical feasibility and failure modes before execution. Integrated within a lookahead planning loop, Learn2Fold enables robust generation of physically valid folding sequences for complex and out-of-distribution patterns, demonstrating that effective spatial intelligence arises from the synergy between symbolic reasoning and grounded physical simulation.
\end{abstract}

\section{Introduction}

Recent advances in generative AI have enabled the synthesis of increasingly complex visual content, including images, videos, and 3D assets~\cite{li2025triposg, chen2023scenedreamer, lu2024direct2, nam20223d, gao2022get3d}. However, most of these successes focus on generating static or perceptual representations, where physical feasibility and execution constraints are either ignored or only weakly enforced. Extending generative models beyond visual plausibility toward physically executable processes remains an open and largely unexplored challenge. This challenge becomes particularly pronounced in tasks that require long-horizon reasoning under strict geometric and topological constraints. While recent progress in deformable object manipulation, such as cloth folding~\cite{tian2025diffusion, liu2025learning, lee2024learning, li2015folding, geminiroboticsteam2025geminiroboticsbringingai}, has demonstrated impressive results, these settings benefit from the inherent compliance and error tolerance of amorphous materials. Garments can accommodate local inaccuracies through smoothing and deformation, allowing learning-based methods to recover from imprecise actions. In contrast, origami folding operates under a fundamentally different regime. Origami is the art of transforming a flat sheet into a three-dimensional structure through a sequence of folds, governed by strict geometric axioms and topological constraints~\cite{5509439, lang2011origami}. A single misplaced crease does not merely introduce a local artifact, but can violate surface topology or render all subsequent folding steps mathematically infeasible. As a result, origami demands precise coordination of discrete topological changes and continuous geometric motions over long horizons, with little tolerance for error.

In this work, we adopt origami folding as a challenging and principled testbed for studying constraint-aware generative planning. Digitally representing and generating origami processes requires modeling both a structured crease pattern and the progressive, constraint-driven folding dynamics that transform a flat sheet into a valid 3D shape. Despite its conceptual simplicity, origami exposes the core limitations of existing generative approaches and serves as a rigorous benchmark for evaluating long-horizon spatial reasoning under hard physical constraints.

Prior work on origami generation can be broadly categorized into learning based methods and optimization based approaches. Generative models, including large language models and vision language models~\cite{openai2024gpt4technicalreport, yang2025qwen3technicalreport, zhang2024visionlanguagemodelsvisiontasks, brohan2023rt2visionlanguageactionmodelstransfer}, are trained on large scale multimodal data such as origami videos, images, and textual instructions. These models can produce descriptive tutorials or high level folding guidance conditioned on text prompts or images. However, they typically fail to generate physically executable origami processes, as they optimize for approximate visual plausibility rather than exact physical feasibility, often hallucinating geometries that appear visually coherent but violate folding constraints. In contrast, traditional optimization based methods~\cite{lang2011origami, tachi2010freeform, he2023fabricfoldinglearningefficientfabric} formulate origami generation as a constrained optimization problem, employing techniques such as circle packing or tuck folding algorithms to mathematically guarantee that a target mesh can be folded from a single sheet. These approaches produce simulation ready, physically grounded crease patterns, but require precise 3D mesh inputs, making them difficult to apply to sparse inputs such as a single image or a text prompt. This raises a key question: can we retain the physical rigor and simulation ready representations of computational origami while leveraging the powerful priors of large language and vision language models to reconstruct executable origami processes from enriching semantic descriptions?

To bridge these gaps, we introduce \textbf{Learn2Fold}, a neuro-symbolic framework that formulates origami folding as constraint-aware program induction. Our key insight is that robust generation requires separating \emph{proposal} from \emph{verification}. Instead of blindly decoding a sequence, Learn2Fold operates in a propose, verify loop. We leverage a Large Language Model (LLM) to propose high-level structured action tokens, utilizing its semantic planning capabilities. However, acknowledging that LLMs lack intrinsic physics grounding, we integrate a learned Graph-Structured World Model for lookahead planning. This world model acts as a differentiable surrogate simulator, allowing the system to \textit{imagine} the geometric consequences of actions and prune branches that lead to invalid states before execution. We propose a symbolic simulator that performs final constraint verification, complementing neural proposal and learned lookahead with exact geometric feasibility checks.

\begin{figure*}[htbp!]
    \centering
    \includegraphics[width=1\linewidth]{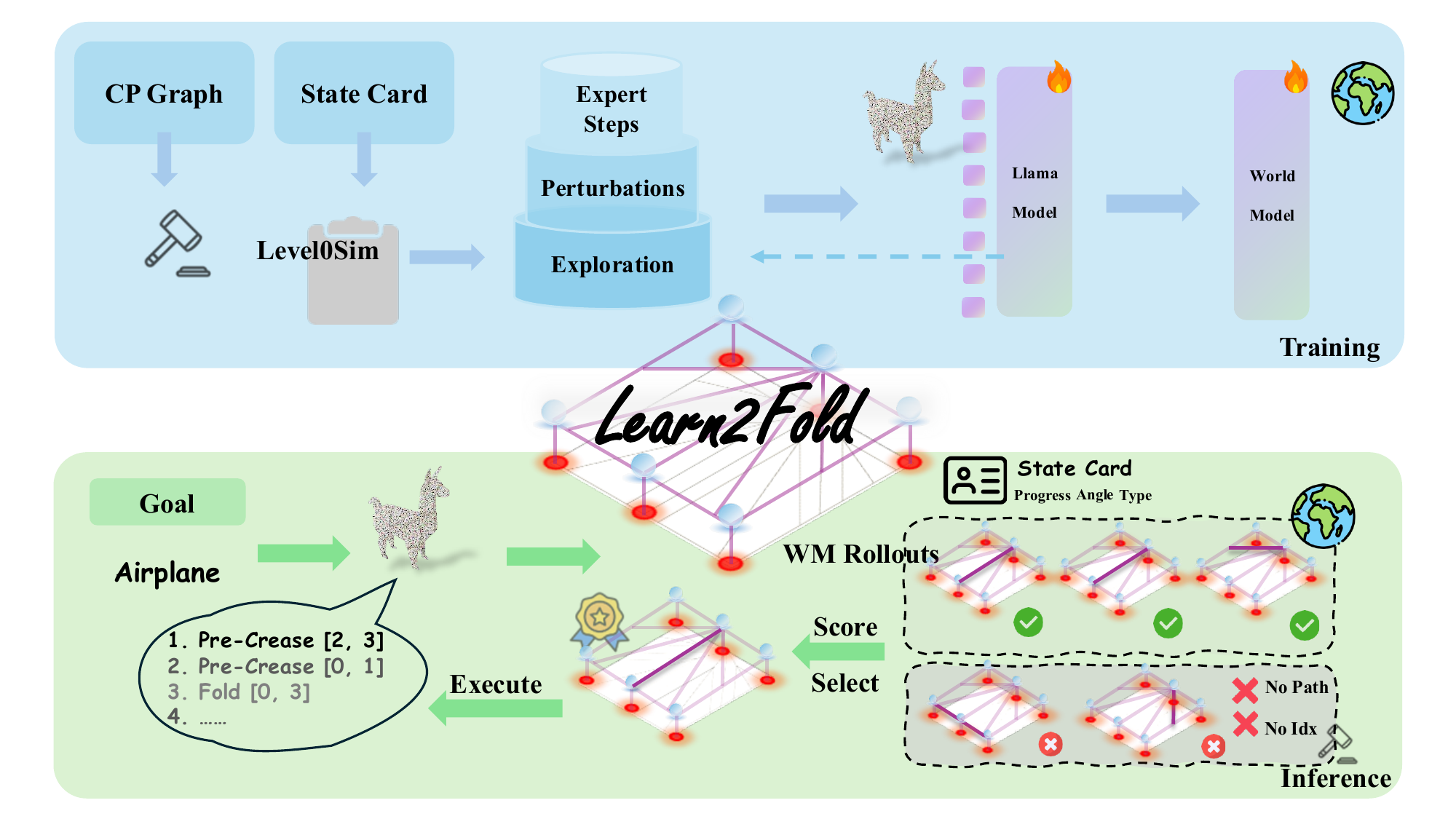}
    \caption{\textbf{Overview of Learn2Fold.}
Learn2Fold formulates origami folding as constraint-aware sequential program generation. During training, a symbolic Level-0 simulator enables scalable data generation and supervision for both a language-based proposal model and a learned world model. At inference time, Learn2Fold combines LM proposals with world-model rollouts and MPC to robustly plan folding sequences under hard constraints.}
    \label{fig:pipeline}
\end{figure*}

Our contributions are summarized as follows:
\begin{itemize}
    \item We propose Learn2Fold, a novel framework for origami process generation that integrates a Large Language Model (LLM) for high-level structured action proposal with a learned Graph-Structured World Model for physics-aware lookahead planning and verification.
    \item A scalable, simulation-driven data curation engine for origami that generates large-scale folding transitions using counterfactual perturbations and propose a new origami dataset, OrigamiCode dataset containing structured folding programs and verified transitions for learning origami folding dynamics.
    \item We validate the effectiveness of the proposed method through comprehensive experiments, demonstrating robust generalization to out-of-distribution physically valid and executable origami generation.
\end{itemize}

\section{Related Work}

\subsection{Structured and Constraint-Aware Generation}

Recent generative models have demonstrated remarkable proficiency in synthesizing high fidelity assets, ranging from static 3D shapes~\cite{wang2023prolificdreamer, voleti2024sv3d, Lan_2025, li2025step1x3dhighfidelitycontrollablegeneration} to dynamic video sequences~\cite{bruce2024geniegenerativeinteractiveenvironments, rombach2022highresolutionimagesynthesislatent, ramesh2022hierarchicaltextconditionalimagegeneration, huang2025vistav2worldimaginationindoor}. However, modeling progressive shape formation processes like origami folding still remains an open challenge. Unlike one-shot generation methods that directly predict a final geometry, origami folding is intrinsically an \textit{executable, long-horizon action sequence}. This process operates on a complex hybrid discrete-continuous state space. This task involves discrete topological changes such as face layering, connectivity updates, coupled with continuous kinematic transformations. Crucially, this generation process is governed by strict physical validity. Every folding step must satisfy hard geometric and topological constraints, such as flat-foldability and self-intersection avoidance; a minor violation in early steps compounds, rendering the final result physically invalid. Consequently, this setting demands structured generation paradigms rather than unstructured end-to-end inference. To address similar structural challenges, recent works have adopted intermediate representations, such as scene graphs or layouts~\cite{johnson2018imagegenerationscenegraphs, xu2017scene, liu2025pixels}, to anchor object relations and reduce spurious outputs. Another line of research integrates constraint-aware decoding or verifier-guided search to ensure validity~\cite{anderson-etal-2017-guided, yan-etal-2021-control, pun2025generating}. For instance, recent structural synthesis models like BrickGPT~\cite{pun2025generating} rely on reactive rollback to filter out physically unstable steps. While these assembly generation systems effectively combine auto-regressive proposals with rollback mechanisms, straightforward backtracking becomes computationally prohibitive for complex folding sequences. Distinguishing our work from these approaches, we propose a CP-grounded folding program equipped with diagnostic feedback. Instead of binary success or failure checks, our model performs causal attribution to identify why a fold failed, enabling efficient planning and recovery even on out-of-distribution crease patterns.

\subsection{Computational Origami}

Origami folding is fundamentally governed by rigorous mathematical rules concerning develop ability and flat-foldability like Kawasaki's and Maekawa's theorems\cite{bern1996flat, demaine2007geometric, hull2002flatfolds}. To simulate these complex behaviors computationally, researchers have developed kinematic models that treat creases as rotational hinges. Early works focused on rigid origami, modeling the mesh as discrete rigid facets connected by joints \cite{tachi2009simulation, tachi2010freeform}. To alleviate this, more recent approaches, such as the bar-and-hinge model used in Origami Simulator \cite{ghassaei2018fast}, introduce compliance to approximate the elastic deformation of paper, enabling real-time folding visualization. While these simulators provide ground truth physics, they are purely forward-process tools where they calculate the geometric consequence of a given fold but do not posses the agency to plan a sequence or reason about hight-level semantic goals.

The problem of generation a crease patter (CP) for a target 3D shape has traditionally been formulated as a geometric optimization problem. Pioneering systems like TreeMaker \cite{lang2011origami} and Origamizer \cite{tachi2009simulation} use circle packing or tuck-folding algorithms to mathematically guarantee that a specific mesh can be folded from a single sheet. However, these methods are strictly geometry-centric and deterministic. They lack the flexibility to handle ambiguous semantic descriptions and are often sensitive to topological errors, where a slight violation in the CP graph renders the entire optimization infeasible. Unlike these optimization-based solvers which require a perfect final mesh as input, our approach treats generation as a sequential decision-making process. This allows for robust recovery from intermediate errors and generalization to out-of-distribution patterns via previous trained structures.

\subsection{World Models}

World models learn action-conditioned dynamics to enable planning via imagined rollouts. This paradigm spans from classical latent-dynamics methods in model-based RL \cite{hafner2019dreamer, hafner2019learninglatentdynamicsplanning, rafailov2020offlinereinforcementlearningimages} to recent foundation-scale video simulators that model physics in rich visual domains \cite{bruce2024geniegenerativeinteractiveenvironments, rigter2024avidadaptingvideodiffusion, huang2025vistav2worldimaginationindoor}. However, pixel-based or latent world models do not directly enforce hard discrete geometric constraints, nor do they naturally produce structured, executable programs. Furthermore, collecting action-labeled interaction data for specialized domains like origami remains prohibitive \cite{doi:10.1126/scirobotics.adt1497}. In our work, we learn a state-level world model over CP-graph states, supervised by scalable synthetic transitions from a deterministic constraint engine. Crucially, our training data includes near-boundary perturbations, exposing the model to both feasible and infeasible outcomes. This learned dynamics model enables efficient model-predictive lookahead, allowing the system to verify action feasibility and recover from proposal errors on out-of-distribution crease patterns.

\begin{figure}[htbp!]
    \centering
    \includegraphics[width=0.5\linewidth]{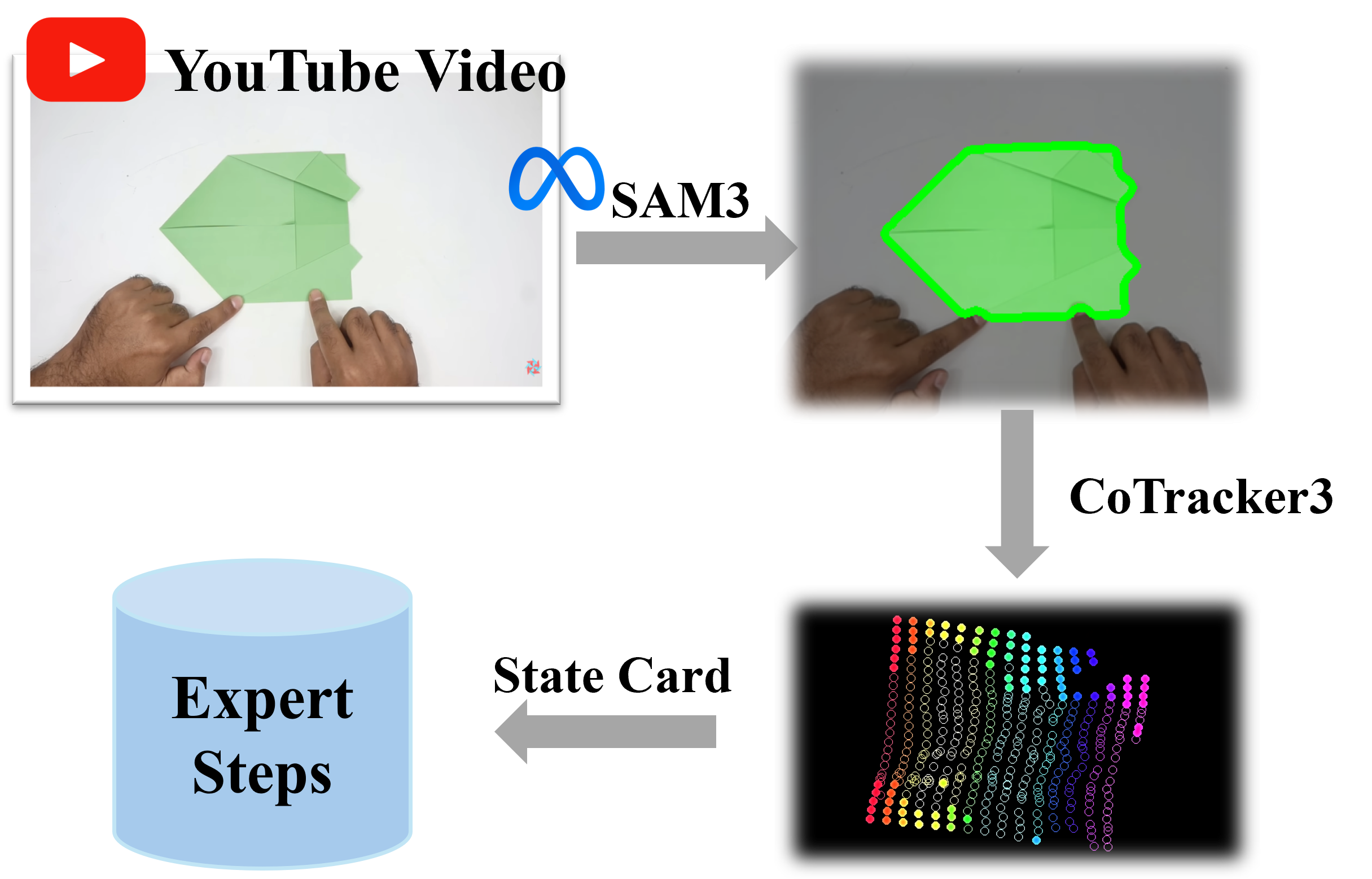}
    \caption{\textbf{Deriving Expert Trajectories from Videos.}
    We show one data source for obtaining expert folding trajectories. In-the-wild instructional videos are processed into State Cards and folding steps, which are then augmented through perturbation and exploration for training.}
    \label{fig:data}
\end{figure}

\section{Method}

We target \textit{physically valid generation} for Computational Origami: at inference time, our agent augments its base proposal policy with a graph-based world model that \textit{imagines} future manifold states and converts them into validity scores for planning (Fig.~\ref{fig:pipeline}).
Our approach, Learn2Fold, tightly couples three components:
\ding{182} a \textit{Canonicalized Graph Representation} that ensures structural invariance;
\ding{183} a \textit{Generative Proposal} Policy that suggests candidate folds based on semantic goals; and
\ding{184} a \textit{Graph-based World Model} that rolls out short-horizon geometric futures.
At test time, we do not only rely on the policy's likelihood; instead, the world model's predictions are fused at the score level via model predictive control (MPC) to rank candidate actions, ensuring strict geometric feasibility without sacrificing generative flexibility.

In the following sections,
Sec.~\ref{sec:data} details the canonicalized state representation.
Sec.~\ref{sec:policy} formalizes the language-conditioned proposal policy.
Sec.~\ref{sec:world_model} introduces the graph world model, which acts as a differentiable surrogate simulator.
Finally, Sec.~\ref{sec:inference} describes the MPC planning strategy that integrates these signals for robust action selection.

\subsection{State Representation and Canonicalization}
\label{sec:data}

We formulate the origami folding process as a sequential manipulation of a graph-structured manifold. An origami instance is represented by a tuple $\mathcal{O}_t = (\mathcal{G}, s_t)$, where $\mathcal{G}$ denotes its static topology and $s_t$ denotes a dynamic state.

\paragraph{Static Graph Topology.}
The crease pattern (CP) is a planar graph $\mathcal{G} = (\mathcal{V}, \mathcal{E})$ with points
$\mathcal{V} = \{v^i \in [0,1]^2\}_{i=1}^{N_v}$ and edges $\mathcal{E}=\{e^j\}_{j=1}^{N_e}$.
Each edge may carry an initial crease type label $z^j_0\in\{\textsc{M},\textsc{V},\textsc{U}\}$ (M: mountain, V: valley, U: unknown).

\paragraph{Canonicalization.}
Raw CP data often contains arbitrary vertex indexing, which hinders learning. To ensure permutation invariance and robust generalization, we apply a deterministic canonicalization process $\Phi: \mathcal{G} \to \mathcal{G}^*$. Specifically, we (i) reindex vertices via lexicographical sorting of coordinates, and (ii) reindex edges based on the sorted endpoint indices. To further eliminate orientation bias, we augment the training data by applying dihedral symmetries (rotations and reflections) to $\mathcal{V}$ prior to canonicalization. This ensures that structurally identical patterns map to the same index space.

\paragraph{Dynamic State.}
We track the folding status using a state vector
$s_t=\big(\alpha_t,\rho_t,z_t,\psi_t,b_t,t\big)$,
where $\alpha_t\in[-\pi,\pi]^{|\mathcal{E}|}$ are signed dihedral angles,
$\rho_t\in[0,1]^{|\mathcal{E}|}$ are progress ratios,
$z_t\in\{\textsc{M},\textsc{V},\textsc{U}\}^{|\mathcal{E}|}$ are crease types,
$\psi_t$ is the global frame angle, $b_t$ is the MV-flip flag, and $t$ is the step counter.

\subsection{Policy Learning via Language Models}
\label{sec:policy}

We frame origami folding as a conditional program induction task. The goal is to learn a policy $\pi_\theta(a_t \mid \mathcal{C}_t)$ that generates a valid folding operation $a_t$ given the current context $\mathcal{C}_t$.

\paragraph{Unified Token Space.}
The action space of folding is inherently hybrid, requiring the selection of discrete graph elements (e.g., target edges) and continuous parameters (e.g., fold angles). To leverage the reasoning capabilities of Transformer-based LLMs, we unify these modalities into a homogeneous vocabulary $\Sigma = \Sigma_{\text{ops}} \cup \Sigma_{\text{graph}} \cup \Sigma_{\text{geo}}$. Continuous geometric parameters are quantized into discrete bins $\Sigma_{\text{geo}}$, while canonicalized graph indices are mapped to semantic tokens $\Sigma_{\text{graph}}$. This formulation transforms the complex control problem into an autoregressive sequence modeling task, enabling the model to capture joint dependencies between topological intent and geometric specifications.

\paragraph{Context and Objective.}
The policy is conditioned on a context $\mathcal{C}_t = (g; \mathcal{G}^*, s_t)$, where $g$ denotes the high-level semantic goal. By operating on the canonicalized graph $\mathcal{G}^*$, the policy learns structure-invariant \textit{folding motifs} (e.g., “rabbit-ear fold”) rather than overfitting to instance-specific identifiers (e.g., vertex indices). We train the model using Maximum Likelihood Estimation (MLE) on expert demonstrations $\mathcal{D}$:
\begin{equation}
\mathcal{L}_{\text{policy}}(\theta) = \mathbb{E}_{(\mathcal{C}, a^*) \sim \mathcal{D}} \left[ -\sum_{k} \log \pi_\theta(a_{t,k} \mid \mathcal{C}_t, a_{t,<k}) \right],
\end{equation}
where $a_{t,k}$ denotes the $k$-th token of the action sequence at step $t$. This supervised pre-training instills the \textit{grammar} of valid folding operations.

\subsection{Graph-Based World Model}
\label{sec:world_model}

While the policy proposes plausible actions, ensuring strict physical feasibility requires rigorous verification. To enable efficient lookahead planning without computationally expensive mesh-based simulations, we learn a differentiable world model $\mathcal{M}_\phi$ that acts as a surrogate simulator.

\paragraph{Residual Graph Dynamics.}
Unlike pixel-based world models~\cite{bruce2024genie} which lack explicit geometric constraints, our model operates directly on the graph state $s_t$. We formulate the transition as a sparse residual update:
\begin{equation}
\begin{aligned}
\Delta\hat{s}_t,\ \hat{m}_t,\ \hat{c}_{t+1}
= \mathcal{M}_\phi(\mathcal{G}^*, s_t, a_t), \\
\hat{s}_{t+1} = s_t + \Delta\hat{s}_t \odot \mathrm{expand}(\hat{m}_t),
\end{aligned}
\end{equation}
where $\hat{m}_t\in[0,1]^{|\mathcal{E}|}$ is a locality mask and
$\hat{c}_{t+1}\in[0,1]^{|\mathcal{E}|}$ estimates per-edge constraint violation likelihood.
$\mathrm{expand}(\cdot)$ broadcasts the per-edge mask to all state channels.

\subsection{Inference via Graph-Guided MPC}
\label{sec:inference}

At test time, we perform a constrained lookahead search on the CP graph $\mathcal{G}=(\mathcal{V},\mathcal{E})$.
At each step $t$, our proposal policy $\pi_\theta$ generates candidate structured actions, which are filtered by a hard verifier (Level-0 simulator) and ranked by the learned world model.

\paragraph{Candidate Sampling.}
We sample $K$ candidate actions from the proposal distribution using nucleus sampling:
\begin{equation}
\mathcal{A}_t = \{a_t^{(k)}\}_{k=1}^{K}, \qquad a_t^{(k)} \sim \pi_\theta(\cdot \mid \mathcal{C}_t).
\end{equation}

\paragraph{Hard Verification (Level-0).}
Each candidate is first evaluated by a deterministic constraint kernel:
\begin{equation}
(\tilde s_{t+1}^{(k)}, v_t^{(k)}, r_t^{(k)}, m_t^{(k)}) =
\textsc{Level0Sim}(\mathcal{G}^*, s_t, a_t^{(k)}),
\end{equation}
where $v_t^{(k)}\in\{0,1\}$ indicates fold validity, $r_t^{(k)}$ denotes the reason for invalidity, and $m_t^{(k)}\in\{0,1\}^{|\mathcal{E}|}$ is the affected-edge mask. We discard invalid candidates and retain
$\mathcal{A}_t^{\text{valid}}=\{a_t^{(k)}\in\mathcal{A}_t \mid v_t^{(k)}=1\}$.

\paragraph{World-Model Rollout.}
For each valid candidate, the world model predicts residual state updates and a soft violation mask:
\begin{equation}
\Delta \hat s_t^{(k)},\ \hat c_{t+1}^{(k)} =
\mathcal{M}_\phi(\mathcal{G}^*, s_t, a_t^{(k)}),\qquad
\hat s_{t+1}^{(k)} = s_t + \Delta \hat s_t^{(k)},
\end{equation}
where $\hat c_{t+1}^{(k)}\in[0,1]^{|\mathcal{E}|}$ estimates per-edge constraint violation likelihood (a soft counterpart of $m_t$).

\paragraph{Action Selection.}
We choose the action maximizing a fused objective of proposal likelihood, goal progress, and feasibility:
\begin{equation}
\begin{aligned}
a_t^* = \arg\max_{a_t^{(k)}\in \mathcal{A}_t^{\text{valid}}}\;&
\frac{1}{|a_t^{(k)}|}\log \pi_\theta(a_t^{(k)} \mid \mathcal{C}_t) \\
&-\lambda_{\text{goal}}\, U_{\text{goal}}(\hat s_{t+1}^{(k)})
+\lambda_{\text{cst}}\, \log\!\Big(\epsilon + 1-\|\hat c_{t+1}^{(k)}\|_\infty\Big).
\end{aligned}
\end{equation}

Here $\lambda_{\text{goal}},\lambda_{\text{cst}}$ balance goal pursuit and constraint satisfaction, and $\epsilon>0$ avoids numerical instability.

\paragraph{Failure and Re-sampling.}
In the case when $\mathcal{A}_t^{\text{valid}}=\emptyset$ or $\max_k J^{(k)}<\tau$, we construct a negative constraint from the predicted violation mask
(e.g., top-$M$ edges with highest $\hat c$) and re-sample candidates under the updated constraint set.

\subsection{Dataset Construction}

\begin{figure*}
    \vspace{-15pt}
    \centering
    \includegraphics[width=1\linewidth]{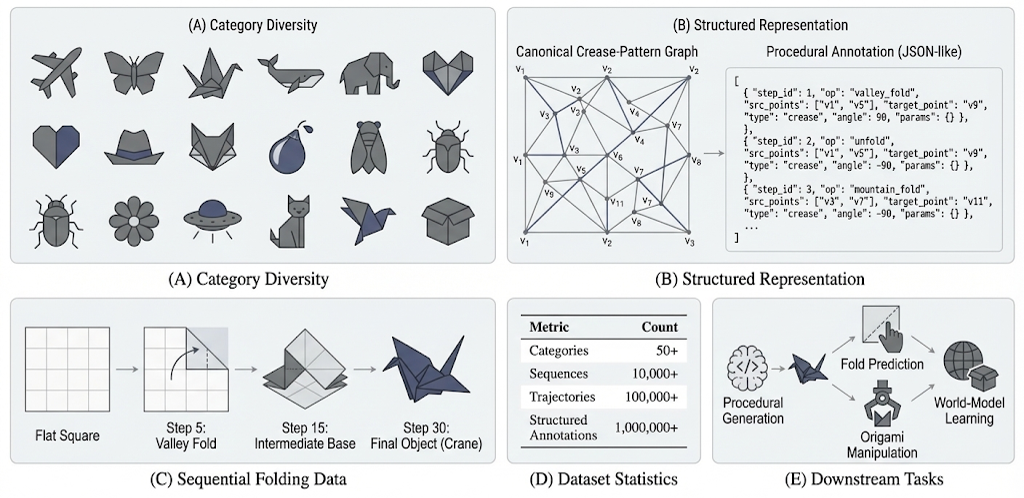}
    \caption{\textbf{Overview of the \datasetname{} Benchmark Dataset.} The dataset features diverse categories, structured representations, sequential folding data, detailed statistics, and benchmark downstream tasks.}
    \label{fig:dataset}
    \vspace{-20pt}
\end{figure*}

\begin{figure}
    \centering
    \includegraphics[width=\linewidth]{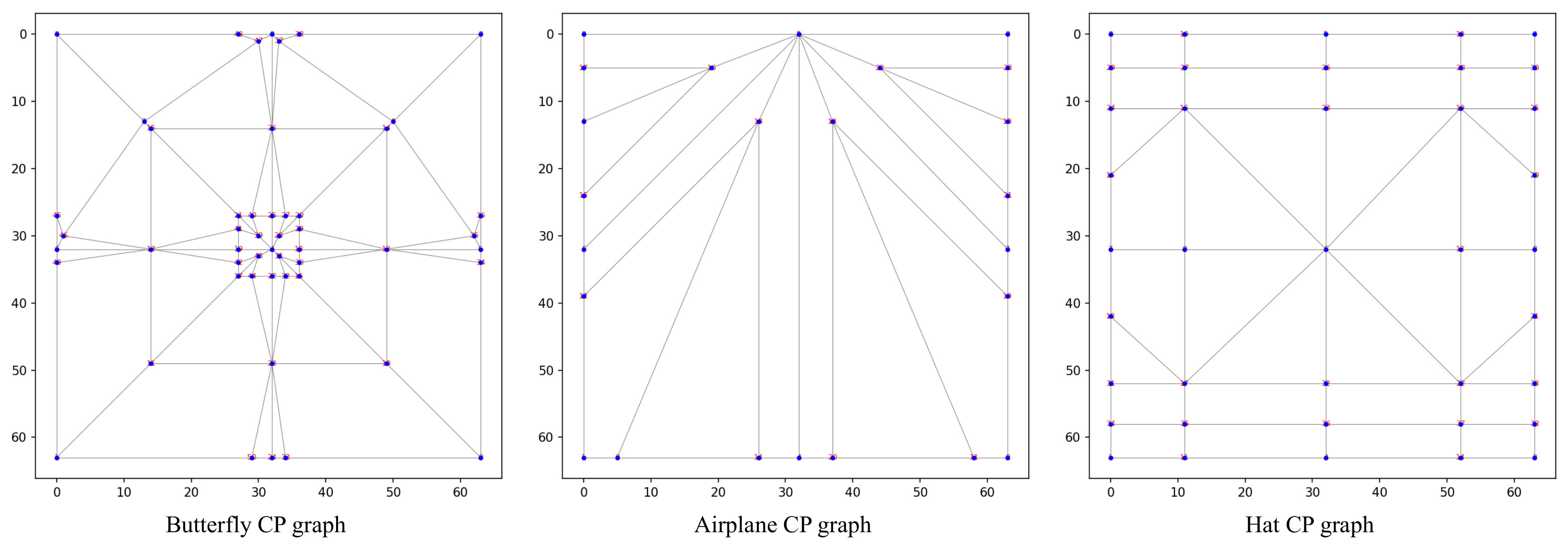}
    \caption{\textbf{Crease Pattern (CP) Graph.} We created CP graph for each case to represent folding sequence.}
    \label{fig:cpgraph}
    \vspace{-10pt}
\end{figure}

To support learning structured origami folding behaviors, we construct the  \datasetname{} dataset (Fig.~\ref{fig:dataset}), a large-scale collection of procedural folding sequences derived from canonical crease pattern (CP) representations. The dataset covers 25 common origami object classes, each defined by a parameterized CP graph (Fig.~\ref{fig:cpgraph}) and an associated folding specification stored in a structured JSON format. Unlike previous datasets dominated by simple shapes, our benchmark is carefully stratified into three difficulty tiers based on step count and non-local dependency: \textbf{Simple (10 categories):} Basic rigid folding structures with minimal layering (e.g., \textit{Airplanes}, \textit{Hearts}, \textit{Cups}).
\textbf{Intermediate (10 categories):} Standard models requiring moderate spatial planning and box-pleating (e.g., \textit{Boats}, \textit{Flowers}).
\textbf{Complex (5 categories):} High-frequency folding sequences with intricate appendage management and strict circle-packing constraints (e.g., \textit{ Cranes}, \textit{Dragons}).

The CP graph encodes the planar crease topology, including vertex connectivity, edge types (mountain or valley), and boundary constraints, while the JSON specification defines the ordered folding operations and geometric parameters required to generate valid intermediate states. Our data curation pipeline derives expert trajectories from instructional sources and further augments them through perturbation and exploration. This process produces a large set of physically consistent folding trajectories, where each sequence records the evolving mesh geometry, crease states, and fold parameters at every step. In total, we collect 5,760 origami process sequences and 75,000 trajectories in the OrigamiCode dataset, which provides structured supervision for learning fold prediction, sequence generation, and origami manipulation tasks.

\section{Experiments}

\subsection{Experiment Setup}

\paragraph{Implementation Details}
We train the world model (WM) using large-scale synthetic folding data generated by the Level-0 simulator.
Specifically, we collect approximately 76,000 transitions through expert demonstrations and constraint-guided perturbations, and train the WM with supervised learning for 50 epochs, which takes about 30 hours on a single NVIDIA RTX Pro 6000 GPU.
The language model (LM) is a lightweight decoder-only transformer fine-tuned to generate structured folding actions under a fixed JSON schema. It is trained using roughly $10^4$ expert folding steps augmented with simulator-verified perturbations, and converges within 6 hours using LoRA adapters on the same hardware.
At inference time, Learn2Fold runs in a model predictive control (MPC) loop, where the LM proposes $N=8$ candidate actions per step, the simulator filters invalid ones, and the WM scores the remaining candidates via short-horizon rollouts to select the final action. All experiments are conducted with fixed random seeds for reproducibility.
\captionsetup[subfigure]{labelformat=empty}
\begin{figure}[t]
    \centering
    \begin{subfigure}[t]{1\linewidth}
        \centering
        \includegraphics[width=\linewidth]{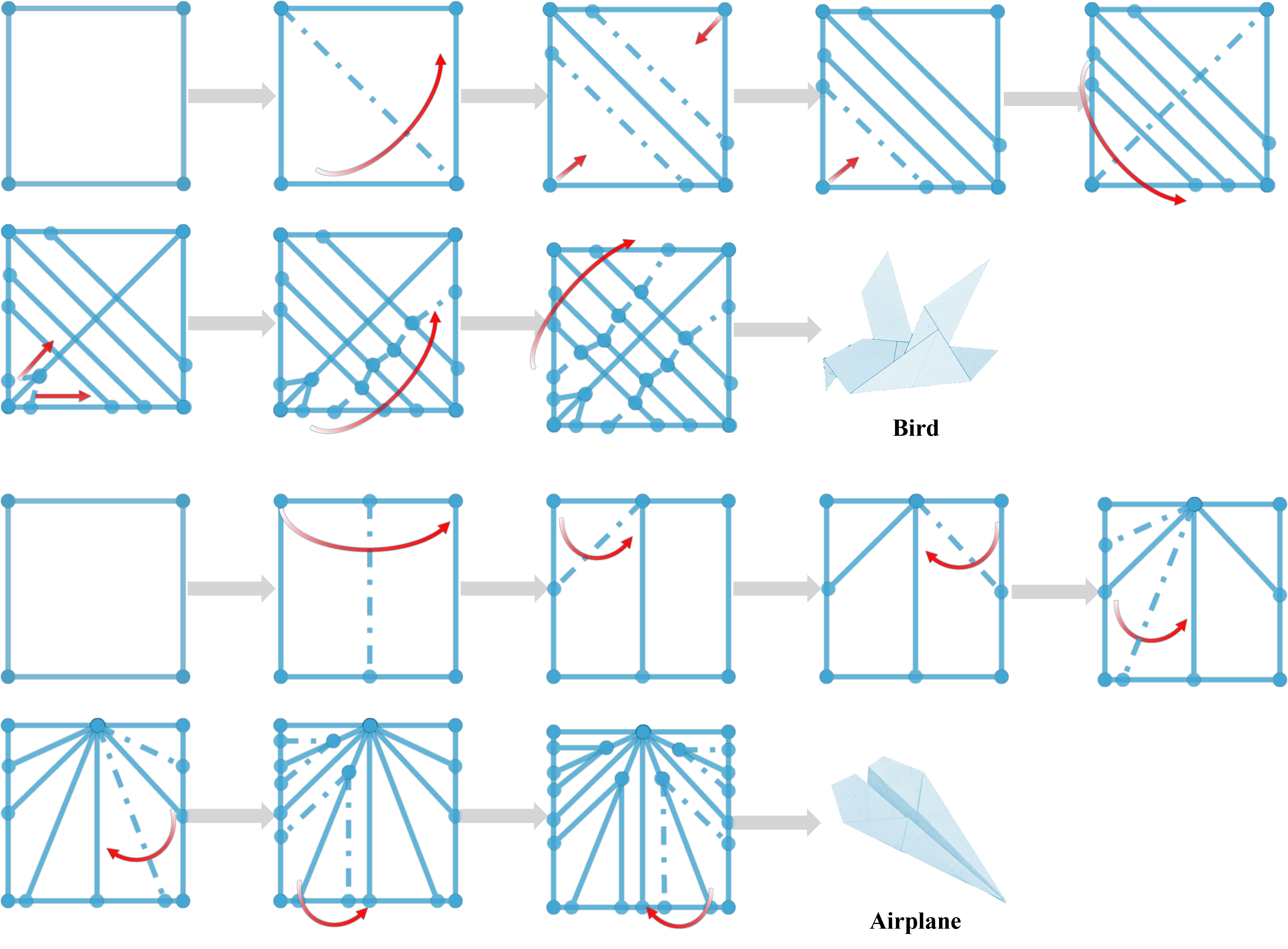}
        \caption{}
    \end{subfigure}

    \vspace{6pt}

    \begin{subfigure}[t]{1\linewidth}
        \centering
        \includegraphics[width=\linewidth]{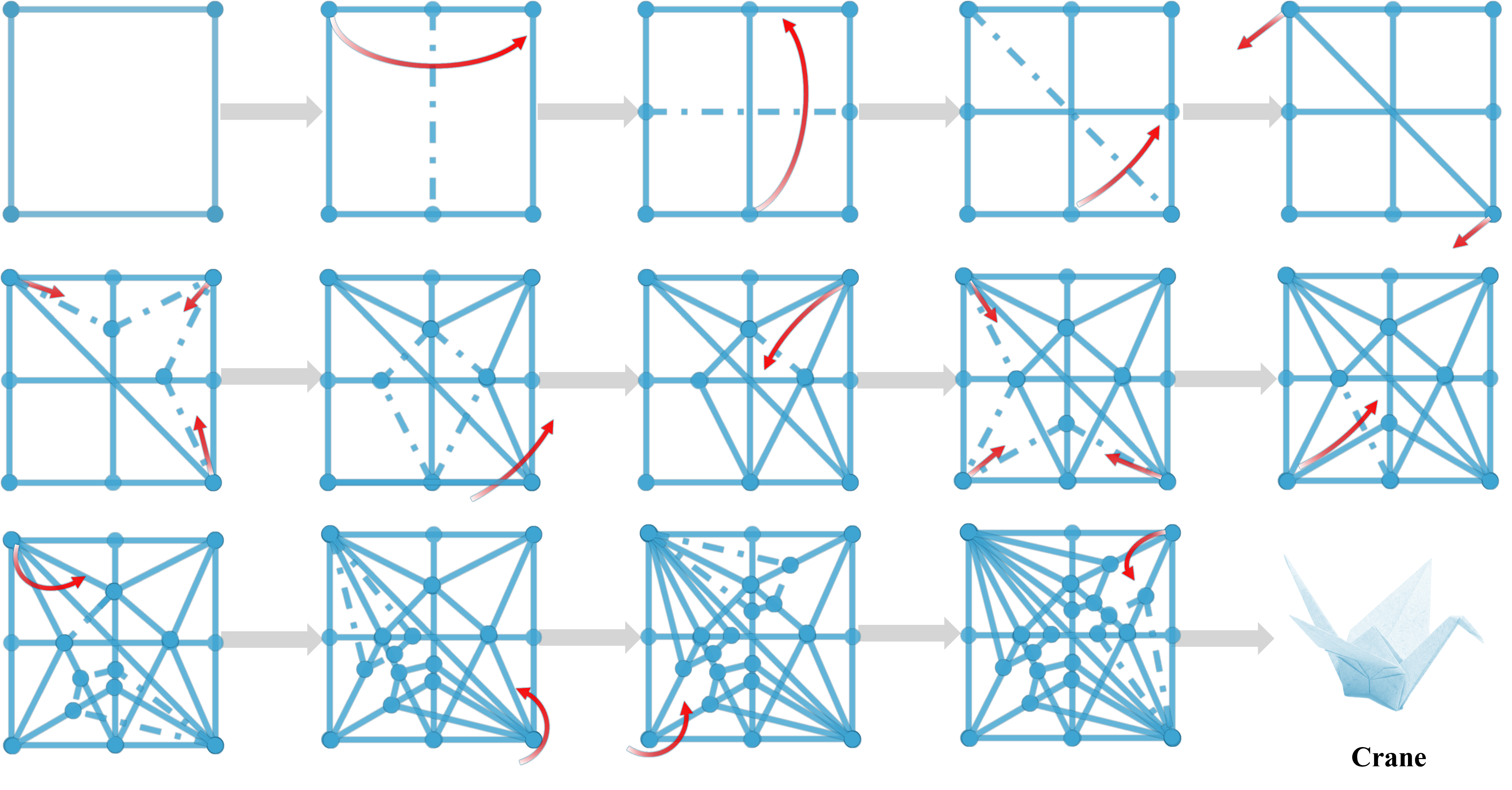}
        \caption{}
    \end{subfigure}

    \caption{\textbf{Learn2Fold results.}}
    \label{fig:results}
\end{figure}

\begin{figure}
    \centering
    \includegraphics[width=1\linewidth]{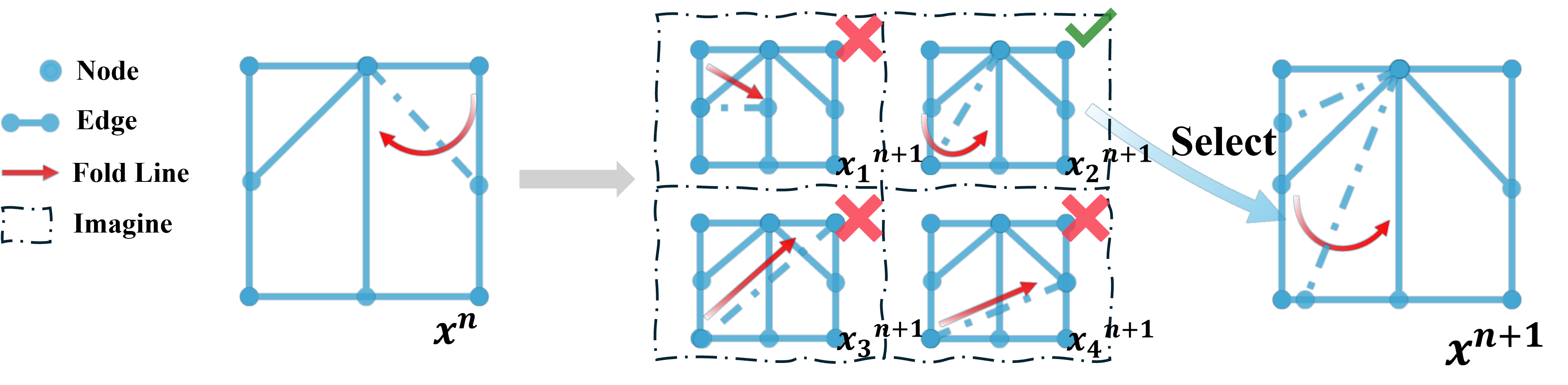}
    \caption{\textbf{Folding with Reasoning.}
    Learn2Fold incrementally constructs origami folding programs in CP-graph space. At each step, multiple candidate actions are evaluated through world-model rollouts, infeasible options are discarded, and the best action is selected for execution, enabling robust folding and recovery under hard constraint}
    \label{fig:statecard}
\end{figure}

\paragraph{Dataset.}
To rigorously evaluate topological generalization, we curate a held-out \datasetname{} benchmark dataset of 25 distinct origami categories that span the full spectrum of folding complexity. This taxonomy allows us to disentangle basic instruction following from complex physical reasoning. Each instance provides a canonicalized CP and a ground-truth program. Following a standard train–test split, 80\% of the data is used for training, while the remaining 20\% is reserved for evaluation.

\paragraph{Baselines.}
We compare our approach against three representative methods. First, we evaluate BrickGPT~\cite{pun2025generating}, a reactive baseline adapted from assembly synthesis that employs a physics-aware rollback mechanism to filter unstable steps through trial-and-error execution and is trained on the proposed OrigamiCode dataset. Second, we benchmark against GPT-5.1 and GPT-5.2, the latest state-of-the-art foundation models. These general-purpose agents are provided with in-context examples to output structured folding programs, representing the upper bound of unconstrained semantic planning without specialized geometric modules. Finally, we compare these against Learn2Fold (Ours), which generates actions under explicit graph-based lookahead verification.

\paragraph{Metrics.}
We evaluate performance at both the step level and the trajectory level. At the step level, we report Precision, Recall, and F1 to measure how accurately each method predicts structured folding actions under a unified action schema, capturing both the correctness and coverage of discrete decisions. At the trajectory level, we report Category Success Rate (Cat-SR) and Edge-IoU to evaluate long-horizon execution performance and structural alignment, respectively. Cat-SR is defined as the fraction of folding sequences that successfully complete the target origami within each category and is macro-averaged across categories to mitigate class imbalance. Edge-IoU measures whether a predicted action affects the correct set of creases by computing the intersection-over-union between the predicted affected-edge set and the simulator-derived ground truth.

\begin{figure*}[t]
    \centering
    \includegraphics[width=1\linewidth]{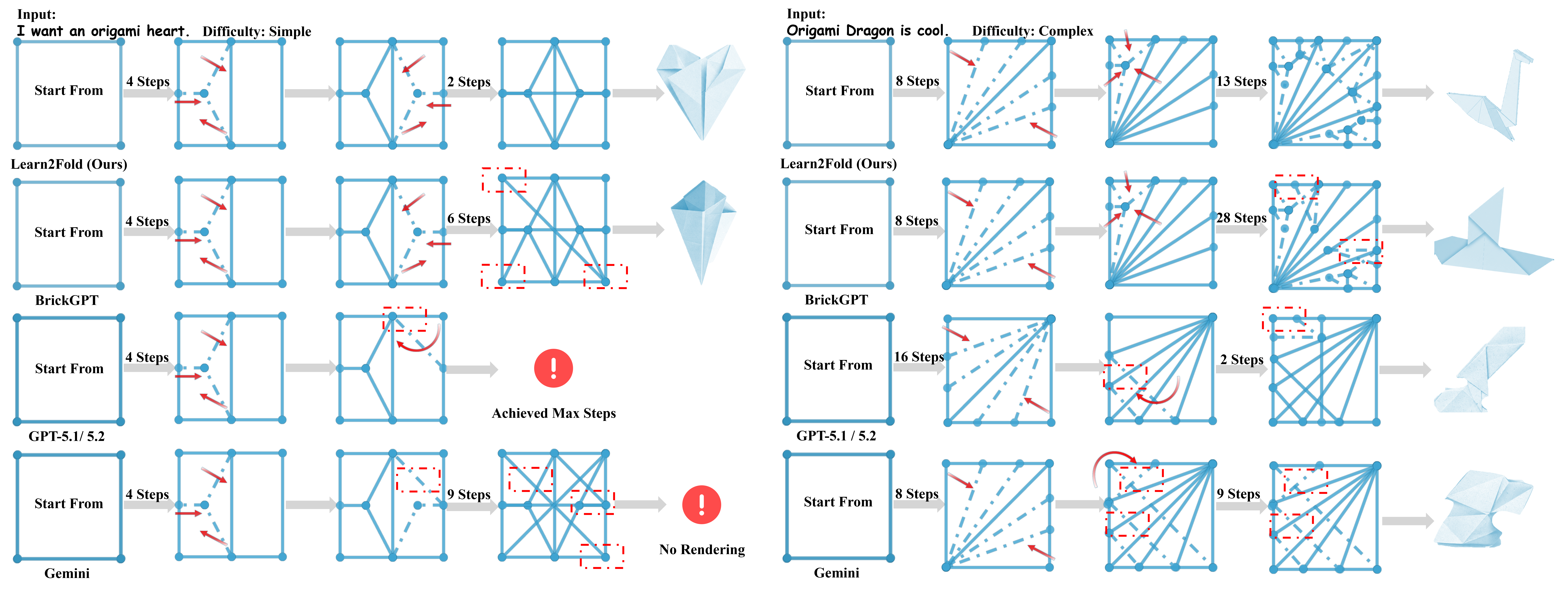}
    \caption{\textbf{Qualitative comparison of folding behaviors across methods.} Learn2Fold produces concise, physically feasible folding trajectories on both simple and complex origami tasks. Baseline methods frequently fail due to invalid actions, early termination, or inability to recover from long-horizon errors, especially on complex crease patterns.}
    \label{fig:qualitative}
\end{figure*}

\subsection{Quantitative Evaluation}
\label{sec:quan}

In the absence of standard benchmarks for origami process generation, following \cite{pun2025generating}, we construct a custom test set comprising 3,840 text prompts spanning 25 categories. From this set, we select 1,150 cases for validation and perform two independent runs per prompt for each method, yielding 7,680 results per method. As shown in Table~\ref{tab:main_compare}, our method outperforms all baselines across all metrics in both step-level accuracy and trajectory-level success. At the step level, Learn2Fold achieves a Precision$_\mu$/Recall$_\mu$/F1$_\mu$ of 0.766/0.711/0.739, substantially exceeding the strongest baseline (GPT-5.1, F1$_\mu$ = 0.266), corresponding to a +47.3 point absolute improvement in F1. In contrast, LLM-based baselines exhibit a pronounced precision–recall imbalance. For example, GPT-5.2 achieves a relatively high Recall$\mu$ (0.358) but very low Precision$\mu$ (0.124), suggesting that while many relevant actions are proposed, they are often imprecise or misaligned with the required structural context since LLMs operate at a coarse semantic level, lacking detailed visual guidance. As a result, they can outline plausible folding intentions but cannot resolve the fine-grained, step-specific details. As for BrickGPT, while it benefits from explicit rollback-based execution and achieves higher precision than LLM-based baselines, it still suffers from limited recall, indicating that its reactive trial-and-error strategy produces coarse and incomplete folding actions and fails to consistently recover the full sequence of required steps.

\begin{table}[t]
\centering
\caption{Main comparison across methods. $^{*}$ indicates prompted models, and $^{\dagger}$ indicates finetuned models.}
\label{tab:main_compare}
\small
\setlength{\tabcolsep}{2.5pt}
\renewcommand{\arraystretch}{1.05}

\begin{tabular}{lccccc}
\toprule
Method & Prec. $\uparrow$ & Rec. $\uparrow$ & F1 $\uparrow$ & Edge-IoU $\uparrow$ & Cat-SR $\uparrow$ \\
\midrule
Gemini\textsuperscript{*} & 0.2874 & \GTwo{0.4213} & \GTwo{0.3420} & 0.1126 & 0.4942 \\
GPT-5.1\textsuperscript{*} & 0.2625 & 0.2996 & 0.2663 & 0.0937 & \GTwo{0.6753} \\
GPT-5.2\textsuperscript{*} & 0.1243 & 0.3575 & 0.1648 & \GTwo{0.1322} & 0.1600 \\
BrickGPT\textsuperscript{$\dagger$} & \GTwo{0.3969} & 0.2250 & 0.2461 & 0.0505 & 0.5455 \\
\midrule
\textbf{Ours} & \GOne{0.7661} & \GOne{0.7113} & \GOne{0.7394} & \GOne{0.5820} & \GOne{0.8912} \\
\bottomrule
\end{tabular}

\vspace{-2mm}
\end{table}

\subsection{Qualitative Study}
Fig.~\ref{fig:qualitative} presents qualitative comparisons between Learn2Fold and baseline methods on representative examples from the same test set described in Sec.~\ref{sec:quan}. LLM-based baselines typically fail after only a few steps. While the initial actions are often semantically plausible, errors quickly accumulate due to the lack of explicit geometric state tracking. As a result, many predicted steps are either structurally incorrect or misaligned with the underlying crease pattern, causing premature termination of the folding process.
BrickGPT exhibits improved stability in the early stages of execution. In several examples, the first few predicted actions (e.g., the first three to four steps) are valid and physically feasible, benefiting from its rollback-based mechanism. However, as folding sequences grow longer, BrickGPT struggles to maintain long-term consistency, as it fails to capture fine-grained dependencies between distant steps, resulting in incorrect or incomplete long-horizon folding processes and limiting its ability to represent detailed step-by-step origami procedures for complex models. While Learn2Fold consistently produces coherent and fine-grained folding sequences across the entire trajectory. By explicitly modeling the folding state and verifying feasibility at each step, Learn2Fold maintains structural consistency and accurately captures the intended origami process, even for long and intricate folding sequences. These qualitative results corroborate the quantitative findings and highlight the advantage of explicit state modeling for reliable origami process generation.

\begin{table}[t]
\centering
\small
\setlength{\tabcolsep}{3pt}
\renewcommand{\arraystretch}{1.05}
\caption{Ablations on IID (top) and OOD (bottom). \TopOne{Blue} and \TopTwo{teal} indicate the best and second-best results.}
\label{tab:main-results}

\begin{tabular}{l ccc}
\toprule
Method & Step Valid $\uparrow$ & Traj SR $\uparrow$ & Goal Dist $\downarrow$ \\
\midrule
LM & \TopOne{70.8\% $\pm$ 45.5\%} & 22.2\% $\pm$ 45.5\% & 0.796 $\pm$ 0.194 \\
LM+WM & 54.2\% $\pm$ 49.8\% & \TopTwo{25.0\% $\pm$ 43.3\%} & \TopTwo{0.759 $\pm$ 0.214} \\
Full (Ours) & \TopTwo{64.2\% $\pm$ 41.8\%} & \TopOne{33.3\% $\pm$ 47.1\%} & \TopOne{0.855 $\pm$ 0.196} \\
\bottomrule
\end{tabular}

\vspace{2mm} 

\begin{tabular}{l ccc}
\toprule
Method & Step Valid $\uparrow$ & Traj SR $\uparrow$ & Goal Dist $\downarrow$ \\
\midrule
LM & \TopOne{47.6\% $\pm$ 29.2\%} & \TopTwo{20.7\% $\pm$ 55.5\%} & 0.633 $\pm$ 0.192 \\
LM+WM & 32.3\% $\pm$ 28.7\% & 17.8\% $\pm$ 51.7\% & \TopTwo{0.560 $\pm$ 0.248} \\
Full (Ours) & \TopTwo{41.2\% $\pm$ 32.3\%} & \TopOne{27.7\% $\pm$ 50.1\%} & \TopOne{0.487 $\pm$ 0.353} \\
\bottomrule
\end{tabular}
\end{table}

\subsection{Ablation}
We conduct an ablation study of the key components of our framework to evaluate the contribution of each proposed component to the final origami process generation's performance. We progressively ablate the system by comparing three configurations: (i) an LLM-only proposer, (ii) LLM augmented with a learned world model (LM+WM), and (iii) the full system Learn2Fold that further incorporates the Level0Sim constraint kernel (LM+WM+Level0Sim).
These variants are evaluated under both in-distribution (IID) and out-of-distribution (OOD) CP holdout settings using step-level validity, trajectory success rate (Traj SR), and final goal distance. Incorporating the world model alters the decision-making behavior by introducing short-horizon lookahead. Compared to the LLM-only baseline, LM+WM exhibits a modest improvement in trajectory-level success (IID Traj SR: 22.2\% $\rightarrow$ 25.0\%), and consistently reduces the final goal distance in both IID (0.796 $\rightarrow$ 0.759) and OOD settings (0.633 $\rightarrow$ 0.560).
However, this improvement comes with a decrease in step-level validity, indicating that the world model prioritizes global progress over local action safety, occasionally selecting actions that are locally risky but potentially beneficial for long-horizon objectives. Adding Level0Sim consistently improves long-horizon performance. The full system achieves the highest trajectory success in both IID and OOD settings, while recovering step-level validity and further reducing final goal distance under distribution shift. Overall, the ablation indicates complementary roles of the LLM proposer, the world model, and Level0Sim, whose combination is required for robust long-horizon folding.

\section{Conclusion}
In this work, we present \textbf{Learn2Fold}, a neuro-symbolic framework for physically valid origami process generation that unifies semantic reasoning with rigorous geometric constraint enforcement. By formulating origami folding as a constraint-aware program induction over a CP graph, Learn2Fold addresses a fundamental limitation of prior generative models: the inability to reliably generate long-horizon, executable action sequences under strict topological and kinematic constraints. We view origami not merely as an application, but as a principled testbed for future spatial reasoning systems, where it exposes core challenges in reasoning over structured space: discrete topological decisions coupled with continuous geometry, irreversible constraints, and long-horizon dependency.

\FloatBarrier
\clearpage

{
    \small
    \bibliographystyle{ieeenat_fullname}
    \bibliography{main}
}

\newpage
\appendix
\section*{\Large Appendix}

\section{More Model and Implementation Details}
\papertitle{} is implemented as a neuro-symbolic planning system over a canonicalized crease-pattern graph representation. Given a high-level semantic goal and the current origami state, the system maintains a unified graph-state interface that is shared across data generation, model training, and inference-time planning. Concretely, each instance is represented by a canonicalized crease-pattern graph together with its dynamic folding state, so that structurally equivalent patterns can be processed in a consistent index space. On top of this representation, we train a lightweight decoder-only language model to generate structured folding actions under a fixed JSON schema, which serves as the proposal policy during inference. To ground the proposal process in physical feasibility, we additionally train a graph-structured world model on large-scale synthetic folding transitions produced by the symbolic Level-0 simulator. The world model is optimized with supervised learning to predict short-horizon state evolution and soft constraint-violation signals, allowing the planner to estimate future failure modes before execution. In our implementation, the world model is trained on approximately 76,000 transitions collected from expert demonstrations and constraint-guided perturbations for 50 epochs, which takes about 30 hours on a single NVIDIA RTX Pro 6000 GPU. The language model is trained on roughly \(10^4\) expert folding steps augmented with simulator-verified perturbations, and converges within 6 hours on the same hardware using LoRA adapters~\cite{hu2022lora}. At inference time, \papertitle{} operates in a model predictive control loop: at each step, the language model proposes \(N=8\) candidate actions, the Level-0 simulator deterministically filters out invalid candidates, and the world model performs short-horizon lookahead scoring over the remaining valid actions to select the final execution step. All experiments are conducted with fixed random seeds for reproducibility.

\section{Evaluation Protocol}

This section provides additional details on the evaluation setup used in the main paper. To assess both structured action prediction and long-horizon folding reliability, we evaluate all methods on a held-out benchmark constructed from 25 origami categories spanning three difficulty tiers: simple, intermediate, and complex. As described in the main paper, the benchmark is designed to cover a broad range of folding complexity, from basic rigid-folding patterns with minimal layering to long-horizon sequences involving non-local dependencies and strict structural constraints. Following the dataset split used throughout the paper, 80\% of the \datasetname{} data is used for training and the remaining 20\% is reserved for evaluation. Each evaluation instance consists of a canonicalized crease pattern together with its target semantic goal and reference folding program. The final test suite contains 3,840 text prompts spanning the 25 categories, from which we select 1,150 cases for validation. For each evaluated method, we perform two independent runs per prompt in order to reduce variance due to stochastic decoding and obtain a more stable estimate of long-horizon generation performance.

We compare \papertitle{} against two classes of baselines. The first class consists of prompted foundation models, including GPT-5.1, GPT-5.2~\cite{singh2025openai}, and Gemini~\cite{team2023gemini}, which are prompted to generate structured folding programs from the same held-out benchmark inputs. These baselines are provided with in-context demonstrations and are evaluated under the same structured output interface as our method whenever possible, so that their predictions can be parsed by the same evaluator. The second class is BrickGPT~\cite{pun2025generating}, a finetuned reactive baseline adapted from assembly synthesis, which uses rollback-based trial-and-error execution to reject unstable steps and is trained on the same \datasetname{} dataset as described in the main paper. This evaluation design allows us to compare semantic-only proposal systems, reactive execution-based baselines, and our full graph-guided planning system under a common benchmark protocol. 

We report both step-level and trajectory-level metrics. At the step level, we measure Precision, Recall, and F1 on structured action tokens under a unified action schema. These metrics quantify how accurately a method predicts the discrete content of each folding action, including operation type and associated structured arguments, while also capturing coverage over the full program. In the main paper, these are reported as the evaluator's category-averaged micro scores. At the trajectory level, we report Category Success Rate (Cat-SR) and Edge-IoU. Cat-SR is defined as the fraction of folding sequences that successfully complete the target origami within each category and is then macro-averaged across categories to mitigate class imbalance. Edge-IoU measures structural alignment between predicted and reference actions by computing the intersection-over-union between the affected-edge set induced by the predicted action and the simulator-derived ground-truth mask. Together, these metrics distinguish local action correctness from global execution success and allow us to separately assess symbolic fidelity, structural grounding, and long-horizon robustness.

For \papertitle{}, evaluation follows the same inference procedure described in the main paper. At each step, the language model proposes multiple candidate actions, the Level-0 simulator discards invalid candidates, and the learned world model scores the remaining valid actions through short-horizon lookahead before one action is selected for execution. A trajectory is rolled out autoregressively until the program terminates or the method fails to produce a valid continuation. For the ablation experiments, we use the same held-out protocol and compare three configurations: LM only, LM+WM, and the full LM+WM+Level0Sim system. These variants are evaluated under both IID and OOD CP holdout settings using step validity, trajectory success rate, and final goal distance, following the setup described in Sec.~4.4 of the main paper. This shared protocol ensures that gains from the world model and the symbolic verifier are measured under the same benchmark conditions rather than under separate task definitions.

Unless otherwise noted, all experiments are conducted with fixed random seeds. The same canonicalized CP representation and unified action schema are used for training, decoding, parsing, and evaluation, which avoids representation mismatch across methods and allows all predictions to be assessed under a common symbolic interface. In the supplementary material, we additionally provide more qualitative examples, representative failure cases, and example structured outputs to further clarify how the above protocol is instantiated in practice.

\end{document}